\documentclass[final, 3p, authoryear]{elsarticle}
\journal{Ecological Indicators} 
\usepackage{amsmath,amssymb,amsthm,amsfonts} 
\usepackage[]{graphicx}
\usepackage{lineno,color} 
\usepackage[dvipsnames]{xcolor}
\usepackage[pdftitle={},
pdfauthor={Stuart R. Borrett},
pdfkeywords={ecological network analysis, network environ analysis,
  centrality, keystone species, foundational species},
pdfpagemode=UseOutlines, bookmarks=T, colorlinks,
linkcolor=blue,citecolor=blue,
pdfstartview=FitH,
urlcolor=red]{hyperref}

\def\tableline{ \vskip .1in \hrule height .6pt \vskip 0.1in}

\def\citeapos#1{\citeauthor{#1}'s (\citeyear{#1})}

\newcommand{\midtilde}{\raisebox{-0.25\baselineskip}{\textasciitilde}}

\definecolor{myBrown}{rgb}{0.7686275,0.6745098,0.4470588}
\definecolor{myGreen}{rgb}{0.7686275,0.8431373,0.5803922}
\definecolor{myPink}{rgb}{0.8549020,0.5882353,0.5803922}

\begin{document} 
\begin{frontmatter}

  \title{Throughflow centrality is a global indicator of the functional
    importance of species in ecosystems}

\author[uncwbio,cms]{Stuart R.~Borrett\corref{cor1}}
\ead{borretts@uncw.edu}

\address[uncwbio]{Department of Biology \& Marine Biology, University
  of North Carolina Wilmington, 601 S.\ College Rd., Wilmington, 28403
  NC, USA}
\address[cms]{Center for Marine Science, University of North Carolina Wilmington}

\cortext[cor1]{Corresponding author. Tel. 910.962.2411; fax: 910.962.4066}

\begin{abstract}
  To better understand and manage complex systems like ecosystems it
  is critical to know the relative contribution of the system
  components to the system function.  Ecologists and social scientists
  have described a diversity of ways that individuals can be
  important; This paper makes two key contributions to this research
  area.  First, it shows that throughflow ($T_j$), the total energy or
  matter entering or exiting a system component, is a global indicator
  of the relative contribution of the component to the whole system
  activity.  Its global because it includes the direct and indirect
  exchanges among community members.  Further, 
  throughflow is a special case of Hubbell status or centrality as
  defined in social science.  This recognition effectively joins the
  concepts, enabling ecologists to use and build on the broader
  centrality research in network science.  Second, I characterize the
  distribution of throughflow in 45 empirically-based trophic
  ecosystem models.  Consistent with theoretical expectations, this
  analysis shows that a small fraction of the system components are
  responsible for the majority of the system activity.  In 73\% of the
  ecosystem models, 20\% or less of the nodes generate 80\% or more of
  the total system throughflow.  Four or fewer nodes are required to
  account for 50\% of the total system activity and are thus defined
  as community dominants.  122 of the 130 dominant nodes in the 45
  ecosystem models could be classified as primary producers, dead
  organic matter, or bacteria.  Thus, throughflow centrality indicates
  the rank power of the ecosystems components and shows the
  concentration of power in the primary production and decomposition
  cycle.  Although these results are specific to ecosystems, these
  techniques build on flow analysis based on economic input-output
  analysis.  Therefore these results should be useful for
  ecosystem ecology, industrial ecology, the study of urban
  metabolism, as well as other domains using input-output analysis.
\end{abstract}

\end{frontmatter}


\linenumbers
\section{Introduction}
Identifying functionally important actors is a critical step in
understanding and managing complex systems, whether it is a fortune 500 company or an
ecosystem. For example, \citet{ibarra1993network} showed that an
employee's power to affect administrative innovation within an
advertising agency was in part determined by their positional
importance within the organization.  
In ecological systems, knowing the relative functional importance of
species or groups of species is essential for conservation biology,
ecosystem management, and understanding the consequences of
biodiversity loss \citep{walker1992biodiversity, lawton94,
  hooper2005effects, jordan06topological, saavedra2011strong}.

Ecologists have several ways of classifying the relative importance of
community members.  \citet{whittaker65} introduced rank--abundance
curves to describe the community richness and indicate the relative
importance of the species, assuming that community importance was
proportional to abundance.  He also presented an alternative
rank--productivity curve that indicated the species importance based
on their net productivity.  Subsequent ecological concepts have built
on this.  Keystone species \citep{paine66, power96} are species whose
importance to the community are disproportionate to their biomass,
like the sea otter in Pacific kelp forests.  Ecological engineers
\citep{lawton94, jones94} are species whose actions create whole new
habitats, such as beavers that transform terrestrial environments into
slow moving aquatic environments.  \citet{dayton72} introduced the more
general term foundational species for fundamentally important species
of many types \citep{ellison2005loss}.  Part of the challenge and the
reason for multiple concepts, is that there are a diversity of ways in
which a species may be important and contribute to a community or
ecosystem.

\begin{figure*}[t]
 \center
 \includegraphics[scale=0.8]{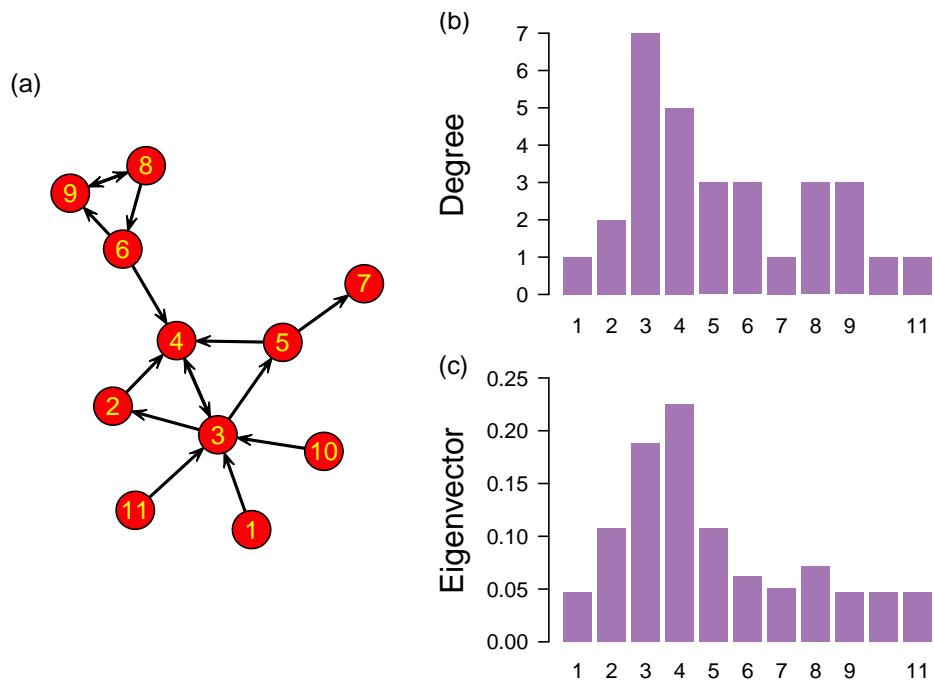}
 \caption{Hypothetical network model (a)  with its associated (b) degree
   and (c) eigenvector centrality.  Degree centrality is a local
   measure while eigenvector centrality is a global indicator of node
   importance.} \label{fig:cent-ex}
\end{figure*}

Faced with the analogous problem of identifying important members of
human communities, social scientists developed the \emph{centrality}
concept \citep[see ][]{wasserman94}.  Centrality embodies the
intuition that some community members are more important, have more
power, or are more central to community function.  Centrality was developed in
the context of network models of communities in which individuals are
represented as nodes of a graph and the graph edges signify a specific
relationship between two individuals such as friendship or
co-authorship (Fig.~\ref{fig:cent-ex}a).  The relationship may or may
not be directed.  Degree centrality is the number of immediately
adjacent neighbors on the graph, and it assumes that more connected
nodes are more central.  It is quantified as the number of edges
incident to the node. In the example graph, node 3 has a degree of 7
(note the separate directed pathways from 3 to 4 and from 4 to 3 shown
as a two headed arrow).  Fig.~\ref{fig:cent-ex}b shows the
distribution of node degrees in the community which indicates that
node 3 is the most central from this local neighborhood perspective.

Some scientists have suggested that the local neighborhood is
insufficient to determine the node's centrality for some applications,
especially exchange networks \citep{hubbell65, bonacich72, estrada10}.
Instead, a node's importance may be increased because one or more of
its neighbors are important.  Network models can capture this
increased neighborhood size by defining a \emph{walk} as a sequence of
edges traveled from one node to another, and walk length ($m$) is the
number of edges crossed.  In the example network, there is a walk from
6 to 2 of length $m=3$ by following $6 \rightarrow 4 \rightarrow 3
\rightarrow 2$.  This enables us to consider the neighborhood $m$
steps aways \citep{estrada10}.  Fig.~\ref{fig:cent-ex}c shows the
eigenvector centrality \citep{bonacich72, bonacich1987power} for the
example network which identifies the equilibrium number of paths
passing through each node as $m \rightarrow \infty$.  In this sense it
is a global centrality measure because it is a ``summary of a node's
participation in the walk structure of the network''
\citep{borgatti05} and captures the importance of indirect as well as
direct interactions \citep{borgatti05, scotti07}.

Degree and eigenvector are only two examples of centrality indicators.
Many centrality measures have been developed and applied in the
literature for complex systems modeled as networks \citep{wasserman94,
  koschutzki05}.  The centrality measures tend to be correlated
\citep{newman06, jordan07,valente08}, but the differences can be
informative \citep{estrada08, baranyi11}.  \citet{borgatti06} provide
a classification of centrality indices and shows how and why different
measures are useful for different applications.


Ecologists have applied the centrality concept in several ways.  For
example, landscape ecologists have used centrality to assess the
connectivity of habitat patches, how this connectivity effects
organism movement, and how habitat loss changes the connectivity
\citep{estrada08, bodin10, baranyi11}.  Community and ecosystem
ecologists have developed and used centrality measures to study how
organisms influence each other in transaction networks
\citep{jordan03, allesina09, fann12_ec}.  \citet{jordan06topological}
argue that mesoscale measures, between local and global centralities,
are most useful for ecosystem studies because the impact of indirect
effects tend to decay rapidly as they radiate through the system. 
Recent work used centrality indicators to determine important species
in communities of mutualists \citep{martin2010centrality, sazima2010}.
Collectively, this work shows how a range of centrality indicators can
be useful for addressing ecological questions.

Here, I identify a new centrality indicator for ecology, termed
throughflow centrality $T_j$.  I first recognize that the throughflow
measure ecosystems ecologists have long calculated \citep{patten76,
  finn76, ulanowicz86} is a global measure of node importance in
generating the total system activity.  Further, I show that this is a
special case of Hubbell's status index centrality \citep{hubbell65}.  
I then apply this measure to 45 trophic ecosystem models drawn from
the literature to test two hypotheses regarding ecosystem
organization. The first hypothesis suggested by both
\citet{whittaker65} and \citet{mills1993keystone} is that communities
are composed of a relatively few dominant species and larger group
that are less central.  The second hypothesis is that in ecosystems
the dominant species/groups are expected to be comprised of primary
producers, decomposers like bacteria, and non-living groups included
in ecosystem models like dead organic matter.  This hypothesis stems
from trophodynamic theory and energetic constraints of food chains
\citep{lindeman42, odum59, jorgensen99_open, wilkinson06}

\section{Theory -- Throughflow is a Centrality Indicator}
A core claim of this paper is that the amount of energy--matter
flowing through each node $j$ in an ecosystem network --- termed node
throughflow ($T_j$) -- is a global centrality indicator of the node's
functional importance.  In fact, this centrality measure is a special
case of \citeapos{hubbell65} status score.  Further, this centrality
indicator is more useful for ecologists and environmental scientists
than the classic eigenvector centrality or the recently introduced
environ centrality \citep{fann12_ec} because (1) it is more intuitive
to calculate, (2) it integrates the transient and equilibrium effects
as flow crosses increasingly longer pathways, and (3) it captures the
effects of environmental inputs (outputs) on the system flows.  This
section provides evidence to support these claims.

\subsection{Flow Analysis}
Flow analysis is a major branch of ecological network analysis (ENA)
\citep{patten76, finn76, ulanowicz86, schramski11}. It is an environmental
application and development of \citeapos{leontief66} macroeconomic
input-output analysis first imported to ecology by
\citet{hannon73}. It traces the movement of energy--matter through the
network of transactions in an ecosystem to characterize the
organization and development of the system.

\subsubsection{Model Definition} \label{sec:model} 

Flow analysis is applied to a network model of energy--matter
exchanges.  The system is modeled as a set of $n$ compartments or
nodes that represent species, species-complexes (i.e., trophic guilds
or functional groups), or non-living components of the system in which
energy--matter is stored.  Nodes are connected by $L$
observed fluxes, termed directed edges or links.  This analysis
requires an estimate of the energy--matter flowing from node $j$ to
$i$ over a given period, $\mathbf{F}_{n\times n}=[f_{ij}]$,
$i,j=1,2,\ldots,n$ (note the column to row orientation).  This flux
can be generated by any process such as feeding (like a food web),
excretion, and death.  As ecosystems are thermodynamically open, there
must also be energy--matter inputs into the system
$\mathbf{z}_{n\times 1}=[z_i]$, and output losses from the system
$\mathbf{y}_{1\times n}=[y_i]$.  In some applications, outputs are
partitioned into respirations and exports to account for differences
in energetic quality, but this is not necessary
in this case.  For other analyses, it is useful when the amount of
energy--matter stored in each node (e.g., biomass) is also reported,
$\mathbf{x}_{n\times 1}=[x_i]$ \citep{fath99_review}.  The necessary
model data $\mathcal{M}$ can be summarized as $\mathcal{M} =
\{\mathbf{F},\mathbf{z},\mathbf{y},\mathbf{x}\}$.

To validly apply flow analysis, the network model must meet two
analytical assumptions.  First, the model must trace a single,
thermodynamically conserved currency such as energy, carbon, or
nitrogen.  Second, the model must be at steady-state for many of the
analyses.  This means that the sum of the energy--matter flowing into
a node equals that exiting the node such that its storage or biomass
is not changing.  \citet{fath07_netconstruction} offer further
suggestions for better ecosystem network model construction.

\subsubsection{Throughflow}
Given this model, we can apply flow analysis.  The technique has a
dual approach.  The \emph{input oriented} analysis pulls the
energy--matter from the boundary outputs and mathematically traces the
pathways (a sequence of $m$ edges) used to generate them all the way
to the boundary inputs. In contrast, the \emph{output oriented}
analysis pushes inputs into the system and follows their
paths through the system to their boundary loss.  This paper focuses
on the output oriented analysis to support the centrality claims for
brevity and clarity; the input perspective provides similar
support.

The first analytical step is to calculate the node
throughflows ($\mathbf{T}_{n\times 1}=[T_j],\: j=1,2,\ldots,n$). \citet{finn76} showed that the input and output oriented
throughflows can be calculated from the initial model information
$\mathcal{M}$ as follows:
\begin{linenomath}
\begin{align}
  T_i^{\textrm{in}}&\equiv \sum_{k=1}^nf_{ik} + z_i \quad (i = 1, 2,
  \ldots,
  n) \textrm{, and}\\
  T_j^{\textrm{out}}&\equiv \sum_{k=1}^nf_{kj} + y_j \quad (j = 1, 2,
  \ldots, n).
\end{align}
\end{linenomath}
At steady state, $[T_i^{\textrm{ in}}] = [T_j^{\textrm{
    out}}] = \mathbf{T}$ and the amount
of energy--matter stored in the node ($x_j$) does not change through
time.  

\citet{finn76} argued that the sum of the node throughflows, called
total system throughflow ($TST=\sum_{j=1}^nT_j$), is a measure of the
activity or size of the ecosystem functioning.  \citet{ulanowicz90}
interpret $T_j$ as the gross production of the compartment.  Thus,
$T_j$ is the contribution of the $j^{th}$ node to the whole system
functioning or productivity. It is in this sense that throughflow is a
centrality measure indicating the relative importance or contribution
of each node.

\begin{figure*}[t]
 \center
 \includegraphics[scale=0.8]{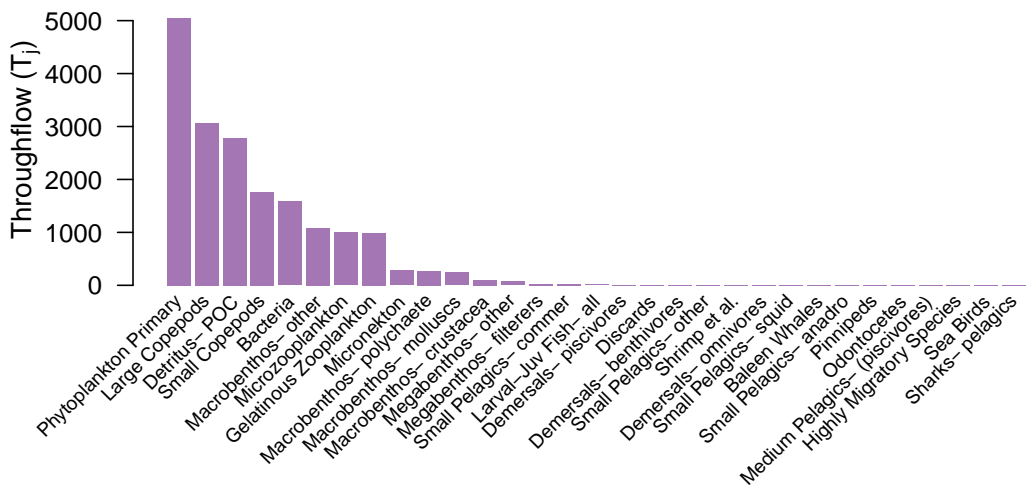}
 \caption{Rank order throughflow centrality for the Gulf of Main
   ecosystem.} \label{fig:gom}
\end{figure*}

Fig.~\ref{fig:gom} shows an example of rank ordered $T_j$ for the Gulf
of Maine ecosystem network \citep{link08}.  This shows the larger
functional importance of phytoplankton, large and small copepods,
detritus, bacteria in this system.  This matches with the theoretical
expectation that primary production and decomposition tend to be the
critical components of ecosystem functioning \citep{wilkinson06}, but
it also points to the importance of smaller consumers in the Gulf of
Maine.  Notice the similarity of this presentation to the
rank--abundance and rank--productivity curves that \citet{whittaker65}
introduced to compare the relative importance of plants in a
community.  Like those original curves, $T_j$ suggests that in this
system there are a few dominant or more important species and a long
tail of functionally less critical species (e.g., Pinnipeds, Beleen
whales, and pelagic sharks).  The application section considers the
generality of both of these patterns.


To facilitate comparisons between centrality measures, it is useful to
consider the node throughflow scaled by the total system throughflow
($T_j/TST$) such that $\sum_{j=1}^n T_j/TST = 1$.  While the
rank-ordering is preserved, rescaling in this way eliminates the units
and differences in total magnitude between systems or other centrality
measures.  This focuses on intensive system differences while ignoring
extensive differences present without the rescaling.  Rescaling
centrality measures is common, though it can introduce its own
challenges \citep{ruhnau00}.

\subsubsection{Path Decomposition}   
Path decomposition of throughflow lies at the core of ENA
\citep{finn76, fath99_review, borrett10_idd}, and shows why $T_j$ is a
global measure of functional importance.  It partitions the flow of
energy--matter from the input (output) over paths of increasing length
(number of directed edges, $m = 0, 1, 2, \ldots, \infty$) within the
system required to generate $T_j$.  Recall that local centrality
measures focus on the connections to a node's nearest neighbors or a
restricted neighborhood, while more global measures consider the
relationships between all nodes within the system.

Path decomposition of flow starts by calculating the output oriented
\emph{direct flow intensities} $\mathbf{G}_{n\times n}=[g_{ij}]$ from
node $j$ to $i$.  These intensities are defined as
\begin{linenomath}
\begin{align} 
g_{ij}&\equiv   f_{ij}/T^{\textrm{out}}_j. \label{eq:G}
\end{align}
\end{linenomath}
Here, $g_{ij}$ is the fraction of output throughflow at donor node $j$
contributed to node $i$.  The $g_{ij}$ values are dimensionless and
the column sums of $\mathbf{G}$ must lie between 0 and 1 with at least
one column less than 1 because of
thermodynamic constraints of the original model
\citep{jorgensen99_open}.

The second step determines the output oriented \emph{integral flow intensities}
$\mathbf{N}=[n_{ij}]$ as
\begin{linenomath}
\begin{align} 
  \mathbf{N} & \equiv \sum_{m=0}^\infty \mathbf{G}^m 
\\ &= \underbrace{\mathbf{I}}_{\textrm{Boundary}} +
  \underbrace{\mathbf{G}^1}_{\textrm{Direct}} + \underbrace{\mathbf{G}^2
    + \ldots + \mathbf{G}^m +
    \ldots}_{\textrm{Indirect}},  \label{eq:N} 
\end{align}
\end{linenomath}
where $\mathbf{I_{n\times n}} = \mathbf{G}^0$ is the matrix
multiplicative identity and the elements of $\mathbf{G}^m$ are the
fractions of boundary flow that travel from node $j$ to $i$ over all
pathways of length $m$.  As the power series must converge given our
initial model definition, the exact values of $\mathbf{N}$ can be
found using the identity $\mathbf{N} = (\mathbf{I}-\mathbf{G})^{-1}$ .
The $n_{ij}$ elements represent the intensity of boundary input that
passes from $j$ to $i$ over all pathways of all lengths.  These values
integrate the boundary, direct, and indirect flows.

We can use $\mathbf{N}$ to recover $\mathbf{T}$ as follows:
\begin{linenomath}
  \begin{align}
    \mathbf{T}&=\mathbf{N}\mathbf{z}.           \label{eq:TN}
  \end{align}
\end{linenomath}
This suggests that the path decomposition of throughflow shown in
equation (\ref{eq:N}) is a true partition of the pathway history of
energy--matter in the system at steady-state.  

The path decomposition in equation (\ref{eq:N}) shows how the
throughflows are a global measure of centrality because the observed
throughflows are generated by energy--matter moving over all pathways
of all lengths such that the whole connected system is considered, not
just a local neighborhood.  Notice that the importance of longer pathways
is naturally discounted as energy--matter is lost as it passes
through nodes in the path.  This discount or decay rate varies among
ecosystems and model types \citep{borrett10_idd}.  Multiplication of
the integral flow matrix by the boundary inputs to recover $T_j$
(equation \ref{eq:TN}) illustrates how the node throughflows capture
the potential effect of heterogeneous boundary inputs known to be a
factor in ecosystems \citep{borrett11_ree}.

\subsection{Hubbell's Status Score}
Before \citet{hannon73} applied \citeapos{leontief65} economic
input--output ideas to ecological systems, \citet{hubbell65} applied
the formalism to social systems.  In doing so, he created a centrality
measure that is known as Hubbell status or Hubbell centrality.
Although Hubbell's initial model was different than the
ecological one presented in section~\ref{sec:model}, the analytical
mathematics is parallel to that shown for throughflow analysis.

\citet{hubbell65} started by modeling the interactions between
individuals in a community using a weighted sociometric choice
matrix $\mathbf{W}=[w_{ij}], (i,j=1,...,n)$, where $w_{ij}$ can be
positive or negative and indicates individual $j$'s indication of the
strength of relationship between him or herself and individual $i$.
The integral relationship strength among the community members
propagated across the whole set of pathways $\mathbf{R}$ are then
determined as
\begin{linenomath}
\begin{align}
\mathbf{R} &= \mathbf{I} + \mathbf{W}^1 + \mathbf{W}^2 + \mathbf{W}^3 + \ldots , \label{eq:Y}
\end{align}
\end{linenomath}
where $\mathbf{I}_{n\times n}$ is again the matrix multiplicative
identity and $\mathbf{W}^m$ is the strength of relationship between
any two community members over paths of length $m$.  When the series
converges, we can find $\mathbf{R}$ exactly as $\mathbf{R} =
(\mathbf{I}-\mathbf{W})^{-1}$.

Building off of this analysis, \citet{hubbell65} defined the
\emph{status score} $\mathbf{S} =[S_i]$ of member $i$ as 
\begin{linenomath}
\begin{align}
  \mathbf{S} &= \mathbf{R\times E} \label{eq:status}
\end{align}
\end{linenomath}
where $\mathbf{E}_{n\times 1} = [e_i]$ are the system exogenous inputs.  


While the initial model was different, the throughflow equation
(\ref{eq:TN}) is identical in form to Hubbell's status shown in
equation (\ref{eq:status}).  Thus, what ecologists call throughflow
$T_j$ is a special case of Hubbell's status index $S_i$ when the model
is defined as in section (\ref{sec:model}).

\subsubsection{Eigenvector and Environ Centrality} \label{sec:EVEC}
To highlight its distinctiveness, $T_j$ is contrasted with two
alternative global centrality measures: eigenvector centrality and
environ centrality. As mentioned in the introduction, eigenvector
centrality (EVC) describes the stable distribution of pathways, or when
weighted as in flow networks the stable distribution of flow, passing
through the nodes \citep{bonacich72, borgatti05}.  In the context of
directed flow networks, \citet{fann12_ec} suggested using the average
of the left $\mathbf{w}$ and right $\mathbf{v}$ hand eigenvector
associated with the dominant eigenvalue of $\mathbf{G}$ to capture
both the input and output, such that
\begin{linenomath}
  \begin{equation}
    EVC = [EVC_i] = \frac{(w_i + v_i)}{2}.  \label{eq:EVC}
  \end{equation}
\end{linenomath}
Note, in this calculation $\mathbf{w}$ and $\mathbf{v}$ are assumed to
have been normalized so that their sum equals 1, which also implies that
$\sum_{i=1}^nEVC_i = 1$.  In symmetric networks like those for which
the eigenvector centrality was first defined $v_i=w_i$ and averaging
is not necessary.  In directed flow networks $v_i\neq w_i$, and EVC
captures the input and output oriented flows intensities.

\citet{fann12_ec} introduced average environ centrality (AEC) and
argued that it is a better centrality indicator for ecosystem flow
networks in part because it captures both the equilibrium dynamics
(like EVC) and transient dynamics that occur along the initial shorter
pathways in equation~(\ref{eq:N}).  This is important because in
highly dissipative systems like trophic ecosystems, a large fraction
of the total transactions might occur in these shorter
pathways. Specifically, \citet{borrett10_idd} found that in nine trophic ecosystem
models 95\% of TST required at most paths of length nine.
AEC is defined as
\begin{linenomath}
  \begin{align}
    \notag EC^{in} &= [ec^{in}_{i}] = \frac{\sum_{j=1}^n n_{ij}}{\sum_{i=1}^n\sum_{j=1}^n n_{ij}}  \\ 
    \notag EC^{out} &= [ec^{out}_{j}] = \frac{\sum_{i=1}^n n_{ij}}{\sum_{i=1}^n\sum_{j=1}^n n_{ij}} \\ 
    AEC &= [aec_{k}] \frac{EC_{j}^{in} + EC_{j}^{out}}{2}. \label{eq:EC}
  \end{align}
\end{linenomath}

Although AEC is an improvement on EVC, both measures still suffer from
two problems.  The first issue is that the calculations required for
EVC and AEC are not intuitive, which could be a barrier to their wider
use in ecology \citep{fawcett2012heavy}.  The second more substantive
issue is that they fail to recognize or capture the external
environmental forcing occurring in these open systems.  Both measures
are built on the non-dimensional flow intensity matrices that
represent the potential flows or the flows if each node had a unit
input.  However, to recover the \emph{realized} or observed system
activity these matrices must be multiplied by the boundary vector as
in equation \ref{eq:TN} \citep[see][]{hubbell65}.  A critical issue
is that the vector of boundary inputs in ecosystem models tends to be
highly heterogeneous \citep{borrett11_ree}, which differentially
excites the potential flow pathways captured in $\mathbf{G}$ and
$\mathbf{N}$.  Given these issues, in many applications $T_j$ is a
better indicator of the functional importance of a node because its
calculation is more intuitive and because it captures the system's
environmental forcing.  

\begin{figure*}[t]
 \center
 \includegraphics[scale=0.8]{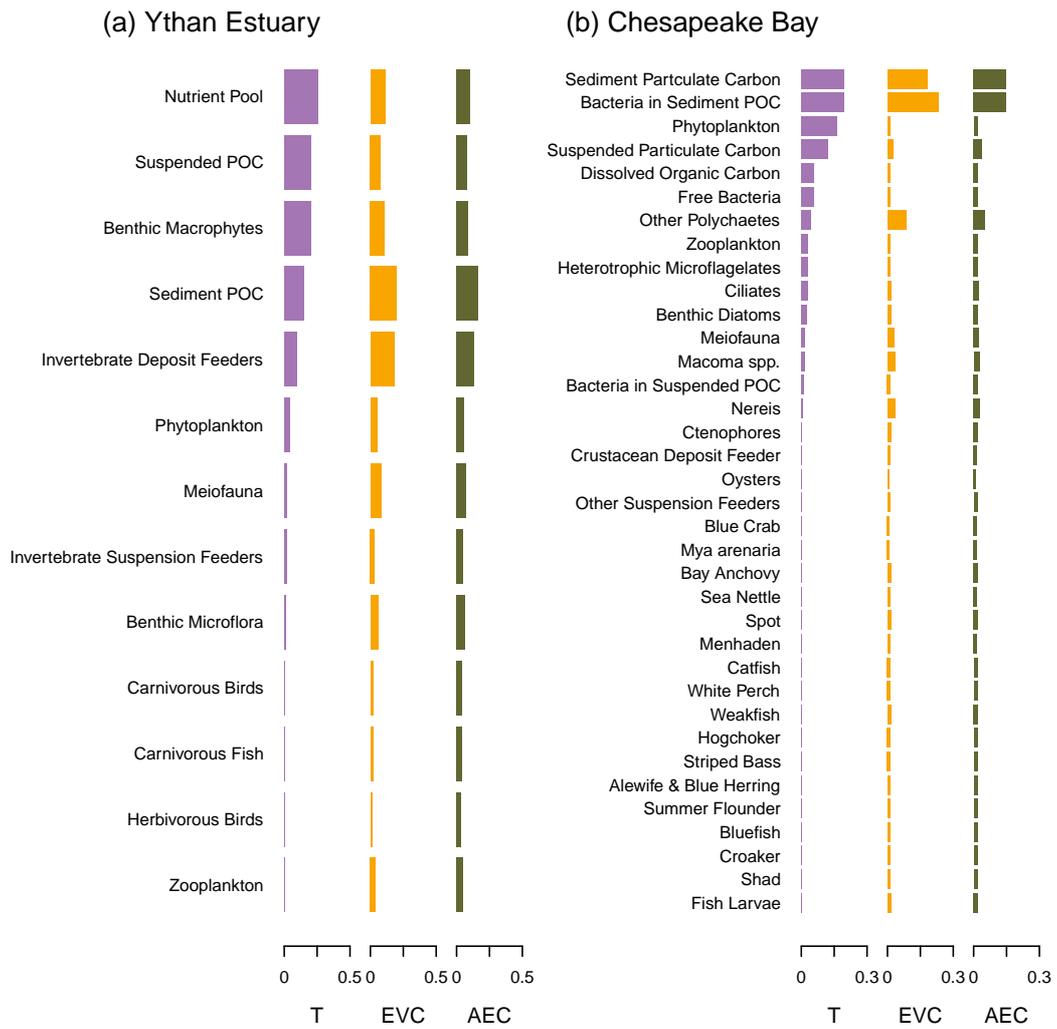}
 \caption{Comparison of throughflow centrality (TC), average eigenvector
   centrality (EVC), and average environ centrality (AEC) in the Ythan
   Estuary (a) and Chesapeake Bay (b) ecosystem networks.  Model
   compartments are rank ordered by TC.} \label{fig:TvEVC}
\end{figure*}

The difference between these indicators can be substantive as
illustrated for the Ythan Estuary and Chesapeake Bay ecosystem models
(Fig.~\ref{fig:TvEVC}).  In the Ythan Estuary, $\mathbf{T}$ is highly
rank correlated with EVC (Spearman's $\rho = 0.79$) and AEC ($\rho =
0.82$), but $\mathbf{T}$ ranks the Nutrient Pool, Suspended POC, and
Benthic Macrophytes as the top three nodes, which is not the case for
the other two indicators.  The first two of these nodes have boundary
input.  The Spearman rank correlation between $\mathbf{T}$ and EVC and
AEC is generally less in the Chesapeake Bay model ($\rho = 0.22$ and
$\rho = 0.55$, respectively).  Again, EVC and AEC discount the
importance of some nodes.  In this case, the top three nodes --
Phytoplankton, Suspended Particulate Carbon, and Dissolved Organic
Carbon -- have non-zero boundary input.  Thus, $\mathbf{T}$ better
captures the importance of nodes that connect the system to its
external environment, and how this influence propagates throughout the
system.

%

In summary, throughflow is a global centrality indicator of the
functional importance of nodes in a flow network.  It is a special
case of what \citet{hubbell65} defined as a status score in sociology.
Due to the natural discounting of longer pathways as energy or matter
dissipates from the system, it has the desirable properties of
mesoscale centrality measures advocated for by \citet{jordan06topological}.
While it is similar to eigenvector and environ centrality measures, it
is more intuitive to calculate and better captures the environmental
forcing of the internal system activity.  The next section applies
$\mathbf{T}$ centrality to characterize the distribution of functional
importance in 45 ecosystem models.

\section{Application --- Materials and Methods}
Given that $T_j$ is a global indicator of an ecosystem component's
functional importance, we can now investigate the distribution of this
importance in ecosystems.

\subsection{Ecosystem Model Database}
I applied flow analysis to 45 trophic ecosystem models selected from
the literature and calculated $T_j$ to investigate the throughflow
centrality distributions (Table~\ref{tab:models}).  To be included in
this data set, the models needed to have at least 10 compartments, have
a food web at their core (i.e., trophic models), and be
empirically-based in the sense that the original modelers were
attempting to represent a real ecosystem and used empirical
measurements to parametrize part of the fluxes.  If two models
existed in the literature for the same system, only the least
aggregated model (higher $n$) was included.  Ten (22\%) of these models are
included in Dr.\ Ulanowicz network collection on his website
(\href{http://www.cbl.umces.edu/~ulan/ntwk/network.html}{http://www.cbl.umces.edu/\midtilde ulan/ntwk/network.html}).
This data set also overlaps 80\% with the models recently analyzed for
resource homogenization \citep{borrett10_hmg}, dominance of indirect
effects \citep{salas11_did}, and environ centrality \citep{fann12_ec}.
The full set of models are available at
\href{http://people.uncw.edu/borretts/research.html}{http://people.uncw.edu/borretts/research.html}.
Forty-four percent of the models were not initially at steady-state,
and were therefore balanced using the AVG2 algorithm \citep{allesina03}.

\subsection{Centrality Comparison}
Rank correlation between $\mathbf{T}$ and AEC and EVC are shown for
the Oyster Reef and Chesapeake Bay ecosystem models in
section~\ref{sec:EVEC}.  Here, this result is
generalized by examining distributions of the Spearman rank
correlation between these measures in all 45 models in our database.

\subsection{Thresholds, and Dominants}
To characterize the $\mathbf{T}$ distributions within a model, I
defined three thresholds. N$_{50}$ is the number of nodes required to
cumulatively account for 50\% of $TST$ when the compartments are rank
ordered based on throughflow (largest to smallest).  If a Monod
function fit the cumulative flow distribution, N$_{50}$ would be
equivalent to the half saturation constant.  N$_{80}$ and N$_{95}$ are
the number of nodes required to recover 80\% and 95\% of $TST$.

\begin{figure*}[t]
\center 
\includegraphics[scale=1]{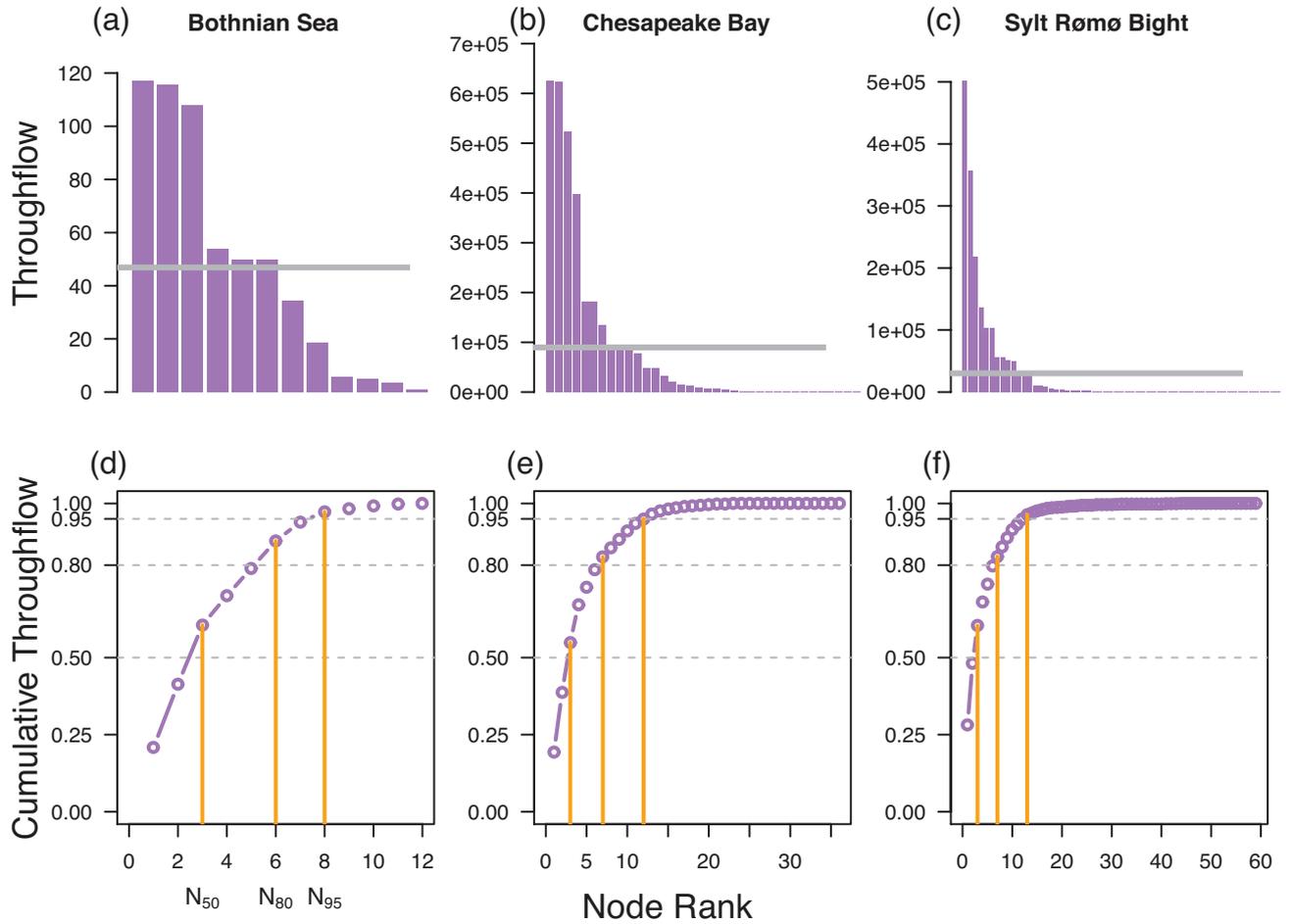}
\caption{Rank ordered throughflow (a, b, c) and cumulative throughflow
  (d, e, f) for the Bothnian Sea (a,d), Chesapeake Bay (b,e), and
  Sylt-R{\o}m{\o} Bight network models (c,f).  Throughflow has the
  units shown in Table~\ref{tab:models}.  The thick horizontal line in
  a, b, and c shows what throughflow would be if each node contributed
  equally.  The vertical lines in (c), (d), and (f) show the nodes at
  which 50\% (N$_{50}$), 80\% (N$_{80}$), and 95\% (N$_{95}$) of the
  total system throughflow is achieved.  For the Bothnian Sea, these
  thresholds are achieved at 3, 6, and 8, respectively.  In the
  Chesapeake Bay they are 3, 6, and 12, while in the Sylt-R{\o}m{\o}
  Bight they are 3, 7, and 13. } \label{fig:cs}
\end{figure*}

These thresholds are illustrated for the Bothnian Sea, Chesapeake Bay,
and Sylt-R{\o}m{\o} Bight ecosystems (Fig.~\ref{fig:cs}).  In the
Bothnian Sea, only three nodes are required to generate 50\% of the
TST (N$_{50} = 3$), while 6 and 8 nodes are required to account for
80\% and 95\% of TST, respectively (N$_{80} = 6$ and N$_{95} = 8$).
In the Chesapeake Bay model, these thresholds were N$_{50} = 3$,
N$_{80} = 6$, and N$_{95} = 12$, and in the Sylt-R{\o}m{\o} Bight they
were N$_{50} = 3$, N$_{80} = 7$, and N$_{95} = 13$.


As the three models shown here have different numbers of compartments,
$n$, it is difficult to compare these thresholds directly.  For better
comparisons, I normalized the thresholds by the model size as
N$_{x}/n*100\%$.  This gives the percent of nodes required to achieve the
$x$\% of $TST$. Fig.~\ref{fig:cs} shows that 33\% of the model nodes
are required to account for 95\% of $TST$ in the Chesapeake Bay model
while only 22\% of the nodes were required in the Sylt-R{\o}m{\o}
Bight model.  This might be interpreted as indicating that system power
is more concentrated in the Sylt-R{\o}m{\o} Bight model.  

There are many ways of defining dominant species or compartments in
ecological systems \citep[e.g.,][]{whittaker65, fann12_ec}.  Here,
dominant compartments in the ecosystem were defined as the smallest
subset of nodes required to recover 50\% of $TST$.  This definition
lets us investigate both how many nodes are required for this ($N_{50}$) as well
as their identity.  For analysis, these compartments were classified
as primary producers (e.g., phytoplankton, submerged vegetation), dead
organic matter (e.g., particulate organic matter, dissolved organic
matter), bacteria (e.g., free living bacteria, bacteria,
benthic bacteria), or other (e.g., filter feeders, meiofauna, large
copepods).  Detritus is technically a mixture of decomposers (some
bacteria) with dead organic matter.  For this analysis, detritus was
grouped with the Dead Organic Matter.    

\section{Results}

\subsection{Centrality Comparison}
\begin{figure*}[t]
\center 
\includegraphics[scale=1]{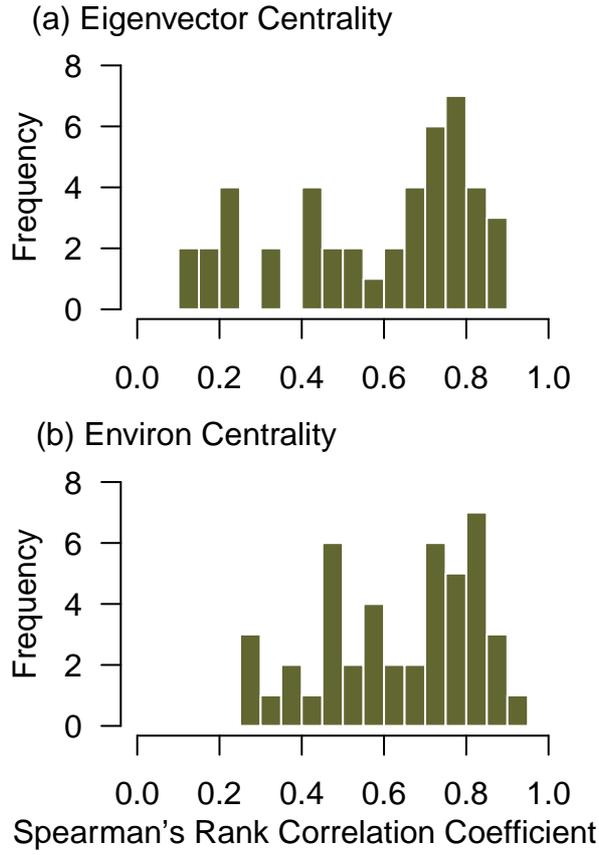}
\caption{Histogram of Spearman rank correlation coefficients between
  throughflow centrality $T_j$ and (a) average eigenvector centrality (EVC) and (b)
  average environ centrality (AEC) in 45 ecosystem models.} \label{fig:CentRankCor}
\end{figure*}

As expected, EVC and AEC tend to be well correlated with $\mathbf{T}$
(Fig.~\ref{fig:CentRankCor}).  The median Spearman rank correlation
between $\mathbf{T}$ and EVC is 0.69, with the values ranging between
0.11 and 0.87.  Throughflow centrality is similarly correlated with
AEC with a median value of 0.69.  The distribution is visibly shifted
to the right and has values ranging from 0.28 to 0.92.  Notice that in
no case is there 100\% agreement or disagreement.

\subsection{Thresholds}
\begin{figure*}[t]
\center
\includegraphics[scale=0.8]{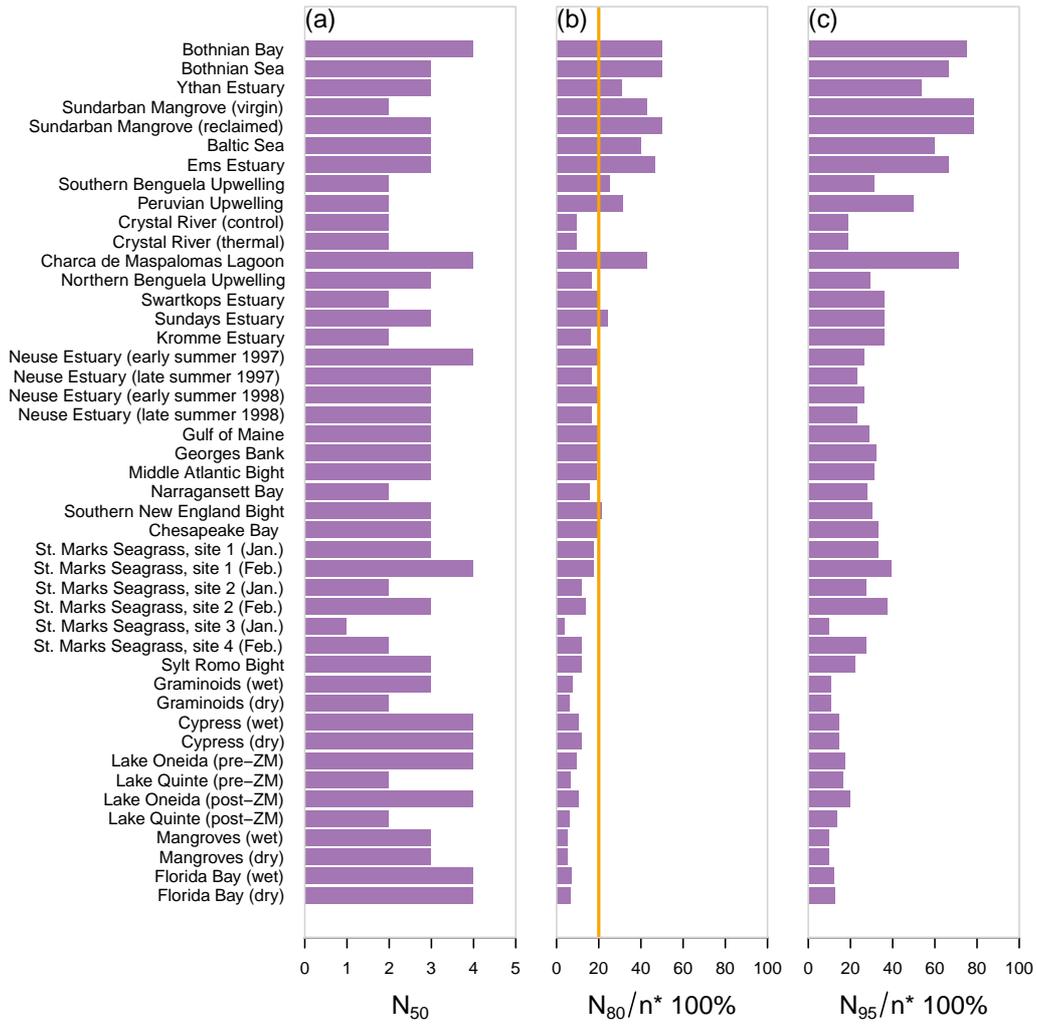}
\caption{Rank order cumulative throughflow thresholds in 44
  empirically based ecosystem models (models ordered by $n$ with
  smallest at the top):  (a) number of nodes required to
  account for 50\% (N$_{50}$), (b) percent of model nodes required to
  achieve 80\% (N$_{80}/n *100\%$), and (c) 95\% (N$_{95}/n *100\%$)
  of total system throughflow.} \label{fig:Nx}
\end{figure*}

Figure~\ref{fig:Nx} shows the cumulative flow development thresholds
($N_{50}$, $N_{80}/n$, $N_{95}/n$) for the 45 trophic network models.
There are several trends to note.  First, the maximum number of nodes
necessary to account for 50\% of $TST$ was 4.  While in the Bothnian
Bay ecosystem model this is 33\% of the nodes, it is only 3.2\% of the
nodes in the Florida Bay model.  Second, as the models increase in
size ($n$) both $N_{80}/n$ and $N_{95}/n$ tend to decline.  Third,
Figure~\ref{fig:Nx}b shows that in the majority (73\%) of the models,
20\% of the nodes or fewer account for 80\% or more of the system
activity.   

\subsection{Dominants}
\begin{figure*}[t]
\center
\includegraphics[scale=1]{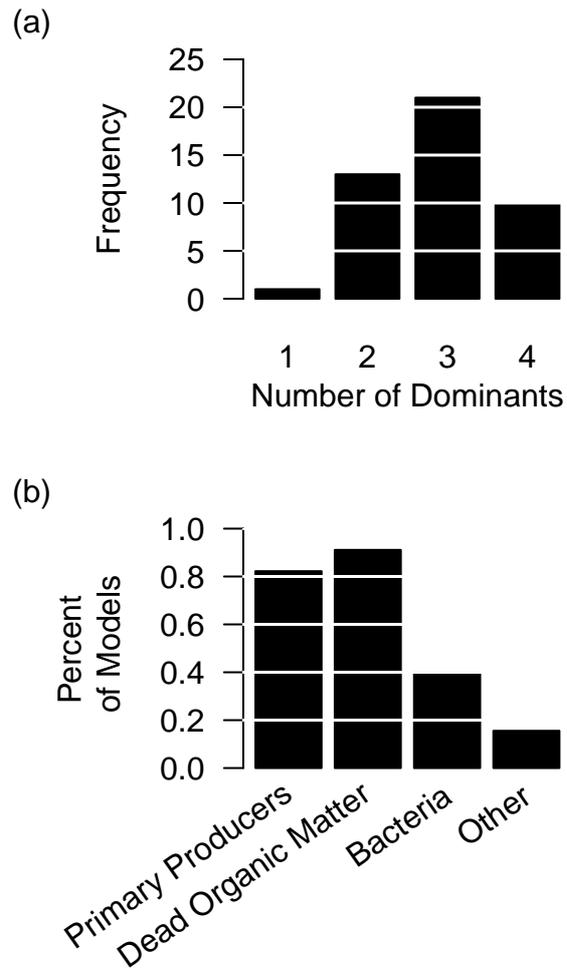}
\caption{Dominant analysis:  (a) the frequency of the 45 models
  with 1, 2, 3, or 4 dominant nodes ($N_{50}$), and (b) the percent of models
  with at least one dominant node in the three functional categories.} \label{fig:doms}
\end{figure*}

Figure~\ref{fig:Nx}a shows that 4 or fewer nodes are required to
account for 50\% of the $TST$ and thus meet the criteria as dominants.
The majority (46\%) of the models analyzed had three dominant nodes,
while another 29\% had only two dominant compartments
(Fig.~\ref{fig:doms}a).

Table~\ref{tab:N50} identifies the 130 dominant nodes in each of the
45 ecosystems. The authors of the original models did not necessarily
use identical categorizations for different ecosystem components, but
it is possible to classify the compartments into four functional
groups: primary producers, dead organic matter, bacteria, and a final
category for anything else (other).  Figure~\ref{fig:doms}b shows the
fraction of models that had at least one dominant in each of these
categories.  Thus, 82\% of the models had at least one dominant
compartment that functioned as a primary producer; 91\% had a dominant
compartment that was categorized as dead organic matter.  Bacteria
were also common. Only 9 of the dominant nodes did not fall into one
of these three categories, and they only appeared in 7 of the models.

\section{Discussion}
Next I consider the theoretical development and its initial ecological
application presented in this paper from three perspectives.  First, I
highlight some of the advantages and disadvantages of recognizing that
system throughflow is a centrality indicator.  Second, I contemplate
the import of this discovery for understanding ecological system
organization, growth, and development. Third, I identify additional
possible applications of this innovation.

\subsection{Throughflow as a Centrality}
A primary contribution of this paper is to recognize that throughflow
$\mathbf{T}$, a measure used by ecologists for some time
\citep[e.g.,][]{finn76, ulanowicz86, fath99_review}, is a centrality
measure as defined in the social science \citep{hubbell65,
  friedkin1991, wasserman94} and now used in general network science
\citep{brandes05}. An advantage of connecting throughflow and
centrality is that ecologists can now access, apply, and further
develop the existing body of work on centrality.  For example, many
centrality measures have been proposed, but sociologists can generally
classify them into one of three types \citep{freeman79, friedkin1991,
  wasserman94, borgatti06}.  The first type are degree based measures.
These measures can vary in the size of the neighborhood considered --
from the immediate local neighborhood to global measures that consider
the whole system \citep[e.g.,][]{estrada10}.  This type of centrality
is generally interpreted as the influence of the node on the network
activity or its power to change the activity
\citep{bonacich1987power}. A second type of centrality is termed
closeness and is based on the shortest paths or geodesic distances
between nodes.  \citet{friedkin1991} suggests that these measures
indicate the immediacy of a node's ability to influence the network.
A third commonly described type of centrality is betweeness
\citep{freeman79, freeman1991centrality}.  A node's betweeness
centrality is its importance in transmitting activity between
individuals or subgroups in the network.  Thus, there is a recognition
of several different but complementary ways in which individuals in a
system can be central.

In this broader context of centralities, Hubbell's status is a global,
weighted, degree based centrality that is typically interpreted as the
node's influence on the whole system activity or its power to change
the whole system activity \citep{borgatti05,brandes05}.  The
formulation allows the node's centrality to be recursively changed by
the centrality of the other nodes in the system as its walk
connectivity is extended.  Although \citet{hubbell65} initially
considered a potentially heterogeneous set of exogenous inputs, in
practice a uniform set of inputs are typically used to consider the
potential centrality.  This is similar to the ``unit'' input
analytical approach often used in network environ analysis
\citep{fath99_review, whipple07, borrett11_ree}.  In the ecological
application of Hubbell centrality, the realized throughflow centrality
is obtained using the observed exogenous inputs.

Ecologists can further benefit from the sociologists previous
applications of centrality.  For example, Hubbell initially used his
centrality as a tool to detect subcommunities or cliques
within the system.  As this is again a common concern for ecologists
\citep{pimm1980food, allesina05_scc, borrett07_jtb}, we may
be able to utilize his procedure to address this problem in the
future.  This would follow \citeapos{krause03} successful application
of a different social network analysis clique finding algorithm to
food webs.    

Another advantage is that we may be able to recognize other ENA
measures as centrality type indicators.  For example, several of
\citeapos{friedkin1991} descriptions of alternative centrality
measures for what he called ``total effects centrality'' were very
similar to what \cite{whipple07} called total environ throughflow
(TET).  Thus, TET may also be a type of weighted degree centrality
measure that indicates the relative contribution of each environ to
the whole system activity.  \citet{hines12} has already begun to
explore this possibility while investigating nitrogen cycling model of
the Cape Fear River estuary.


There are two potential disadvantages of recognizing throughflow as a
centrality indicator.  First, it could contribute to the proliferation
of centrality measures that can be overwhelming.  This has led to
multiple papers trying to identify the unique contributions of
specific indicators amongst a set of competing indicators \citep[e.g.,
][]{newman06, jordan07, valente08, bauer2010, baranyi11}.  In this
case, however, I argue that we are not creating a new centrality index
to add to the confusion, but identifying that a commonly calculated
measure is a form of an existing centrality measure.  A second
disadvantage might be that the current use and implementation of
Hubbell's centrality available in software packages may be simplified
from its original formulation, as appears to be the case in Ucinet
\citep{borgatti02}.  The output of the Hubbell centrality analysis in
Ucinet does not match the throughflow vector as calculated with NEA.m
\citep{fath06}

As expected, $\mathbf{T}$ generally correlates well with average
eigenvector centrality (EVC) and average environ centrality (AEC) for
the 45 models examined.  This suggests that these different global
degree-based centrality measures capture some of the same information
about the relative importance of the nodes for the system function.
However, the correlations were variable -- in some cases the rankings
were quite different (e.g., median Spearman correlations were 0.69 and
the lowest was 0.11) suggesting that each measure captured some unique
information.  Examining both the formulation of the three centrality
measures as well as the example in Figure~\ref{fig:TvEVC}, a key
difference is that $\mathbf{T}$ captures the importance of a node for
connecting the system to the external world.  For example in the Ythan
estuary model, the Nutrient Pool and Suspended POC both have large
inputs that contribute to their importance in $\mathbf{T}$.  Thus in
applications where the boundary inputs are an important consideration,
an indicator like throughflow centrality may be the best choice.  For
example, \citet{borrett11_ree} argued that this system--environment
coupling is critical for ecologists and environmental scientists even
when the analytical focus is on the within system environments.




\subsection{Throughflow and Ecosystem Organization and Development}  

Ecologists have a long interest in the organization, growth, and
development of ecosystems \citep[e.g.,][]{odum69, ulanowicz86,
  jorgensen00, gunderson02, loreau10}.  What are the
processes that create, constrain, and sustain ecological systems?
Scientists investigating this problem have hypothesized a number of
goal functions or orientors that might guide the growth and
development of these self-organizing systems \citep{schneider94,
  muller98, jorgensen07_newecology}.  Hypothesized orientors include
the tendency for ecosystems to maximize power \citep{lotka22,
  odum1955}, maximize biomass or storage \citep{jorgensen79}, maximize
dissipation \citep{schneider94}, and maximize emergy
\citep{odum88}. \citet{fath01} used the network framework to show how
these different orientors can be complementary.

\citet{patten95} suggested that throughflow in network models of
energy flux can be interpreted as a measure of \emph{power} in a
thermodynamic sense.  He argued that $TST$ indicates the total power
output of an ecological system.  This operationalized
\citeapos{lotka22} maximum power principle for evolutionary systems
and \citeapos{odum1955} hypothesis that ecological systems tend to
maximize their power in a network context.  Given this interpretation
of $TST$, $T_j$ is therefore the partial power of each node ($j = 1,2,
\ldots, n$) in the
network. 
Interestingly, this thermodynamic interpretation to throughflow
aligns with the social interpretation of this type of centrality as
the power to influence the system \citep{bonacich1987power}.

Recognizing that network nodes in ecosystem models represent
subsystems in a hierarchical context \citep{allen1982}, then we can
extend the maximum power hypothesis to each node.  As all nodes would
experience the same attraction to increase $T_j$, we might expect the
$T_j$s to be more similar (towards a uniform distribution).  However,
this maximization remains restrained by the evolutionary constraints
of the individual organisms, including their participation within the
existing ecosystem \citep{walsh2009, guimaraes2011evolution}.  For
example, \citet{ulanowicz97, ulanowicz09_auto} argues that the
formation of autocatalytic cycles can be an agency for ecosystem
growth and development.  These cycles can provide the positive
feedback and selective pressure for individual nodes to tend to
increase their $T_j$.  They also provide a selection pressure such
that alternative nodes within an autocatalytic cycle compete for
participation in throughflow and can be replaced by higher performing
entities.  \citet{ulanowicz97} further argues that the tendency of
these cycles for centripitality -- in this context attracting and
capturing more resources -- leads to the emergence of a system
autonomy from the material cause of the system.  Thus, evolutionary
constraints on species and the system constraints of interacting
autocatalytic cycles might increase the variability of $T_j$ despite
the homogenizing effect of the tendency to maximize throughflow.

The throughflow threshold analysis of the 45 ecosystem models
presented here indicates that throughflow centrality is far from
uniform as it appears to follow something more like Pareto's 80-20
rule in which 80\% of the activity is done by 20\% of the group
\citep{reed2001pareto}.  This suggests that throughflow centrality may
be similar to if not exactly the scale free degree distributions
commonly found in other types of complex systems
\citep{barabasi02}. In addition, all but 8 of the dominant or most
central nodes could be classified as primary producers, dead organic
matter, or bacteria.  This aligns with what we might expect from
ecosystem theory in general and the importance of autocatalytic
hypercycles like the autotroph $\leftrightarrow$ decomposer cycle
\citep{ulanowicz97, wilkinson06}.

\subsection{Applications}
Network modeling and analysis, Input-Output Analysis, and material
flow analysis have broad application.  The ideas originated in macro
economics \citep{leontief66} and as has been discussed are used in
both sociology and ecology.  Thus, throughflow centrality may be
useful in multiple domains of inquiry.  

Beyond the theoretical considerations for ecosystem growth and
development, there are a number of ways in which the throughflow
centrality indicator could be usefully applied for ecosystem
management, conservation, and restoration. 
For example, the throughflow centrality analysis
suggests which species or groups of species should be targeted in the
goal is to increase or decrease the system activity.  The impact of
manipulating a more central node should be greater than modifying a
less central node.  

Materials flow analysis is an important tool for industrial ecology
\citep{bailey2004applyingI, bailey2004applyingII, suh2005industrial,
  gondkar2012methodology} and urban metabolism
\citep{kennedy2011study, zhang12, chen12}.  The specific ENA methods
described in this paper have been used to analyze the sustainability
of urban metabolisms \citep{bodini02, zhang10, chen12}.
\citet{chen12} shows how throughflow can be grouped according to
compartment ``trophic levels'' to build productivity pyramids for
cities that are then comparable to expected trophic productivity
pyramids in ecology.  Thus, the recognition that $\mathbf{T}$ is a
centrality indicator could have a broad utility for these disciplines.

ENA is an ecoinformatic tool and shares many goals and characteristics
with network analysis in the field of Systems Biology.  For
example, \citet{hahn2005comparative} showed that genes with higher
centrality tend to be functionally more important in protein-protein
interaction networks.  While thermodynamically conserved flows are not
normally the focus of the systems biology network models (omics)
making it difficult to apply the flow analysis and ENA more broadly,
\citet{kritz2010} suggest a way of liking a metabolic network model to
the underlying chemical fluxes and reactions.  If this technique
proves robust, then the throughflow centrality might be useful in this
domain as well.

\section{Conclusions}

In summary, this paper makes two primary contributions.  First, I show
that throughflow ($T_j$) in network input-output models is a global
indicator of the relative importance or power of each node in the
network with respect the whole system activity.  As calculated in
ecological network analysis, this is a special case of Hubbell
centrality \citep{hubbell65}.  Second, when applied to trophic network
models of ecosystems, throughflow centrality shows the tendency of
this power to be concentrated in a small set of nodes that tend to
categorized as primary producers, dead organic material, or bacteria.
This is consistent with previous theory regarding the growth and
development of ecological systems.

To address the wicked problems \citep{rittel1973dilemmas} of our time like
economic challenges and global climate change, we will need to be both
creative and innovative.  An innovation in this paper is to join the
throughflow concept in flow analysis and the centrality concept
developed in the social sciences.  I expect this to be a useful union
that will enable new analysis and management of complex systems of many kinds
including urban metabolisms, industrial ecosystems, and biogeochemical
cycling and trophic dynamics in natural ecosystems.

\section{Acknowledgments}
Early ideas for this paper were first presented at the 2011 meeting of
the International Society for Ecological Modelling in Beijing, China
and benefited from conversations with several colleagues including
Ursula Scharler, John Schramski, Bernie Patten, Sven J{\o}rgensen, and
Jeff Johnson.  I also appreciate many colleagues sharing their network
models with the Systems Ecology and Ecoinformatics Laboratory,
including Dr. Baird, Link, Ulanowicz, and Sharler.  Further,
D.E. Hines provided comments on early manuscript drafts.
This work was supported in part by UNCW and NSF (DEB-1020944).


\clearpage


\begin{table*}
  \caption{Forty-five empirically-based trophic ecosystem network
    models.} \label{tab:models}
  \tableline 
\begin{center}
\begin{footnotesize}
\begin{tabular}{l l c c r r c r}
Model & units & $n^\dagger$ & $C^\dagger$ & $Boundary^\dagger$ & $TST^\dagger$ & $FCI^\dagger$ & Source \\
\hline \\[-1.5ex]
Bothnian Bay  &  gC m$^{-2}$ yr$^{-1}$  & 12 & 0.22 & 44 & 184 & 0.23 &   \citet{sandberg00} \\
Bothnian Sea  &  gC m$^{-2}$ yr$^{-1}$  & 12 & 0.24 & 117 & 562 & 0.31 &   \citet{sandberg00} \\
Ythan Estuary  &  gC m$^{-2}$ yr$^{-1}$  & 13 & 0.23 & 1,259 & 4,182 & 0.24 &  \citet{baird81} \\
Sundarban Mangrove (virgin) &  kcal m$^{-2}$ yr$^{-1}$ & 14 & 0.22 & 117,959 & 441,214 & 0.16 &  \citet{ray08} \\
Sundarban Mangrove (reclaimed) &  kcal m$^{-2}$ yr$^{-1}$ & 14 & 0.22 & 38,485 & 103,057 & 0.05 &  \citet{ray08} \\
Baltic Sea  &  mgC m$^{-2}$ d$^{-1}$  & 15 & 0.17 & 603 & 1,974 & 0.13 &   \citet{baird91} \\
Ems Estuary  &  mgC m$^{-2}$ d$^{-1}$  & 15 & 0.19 & 283 & 1,067 & 0.32 &  \citet{baird91} \\
Southern Benguela Upwelling  &  mgC m$^{-2}$ d$^{-1}$  & 16 & 0.23 & 715 & 2,546 & 0.31 & \citet{baird91} \\
Peruvian Upwelling  &  mgC m$^{-2}$ d$^{-1}$  & 16 & 0.22 & 14,928 & 33,491 & 0.04 &  \citet{baird91} \\
Crystal River (control)  &  mgC m$^{-2}$ d$^{-1}$  & 21 & 0.19 & 7,358 & 15,063 & 0.07 &  \citet{ulanowicz86} \\
Crystal River (thermal)  &  mgC m$^{-2}$ d$^{-1}$  & 21 & 0.14 & 6,018 & 12,032 & 0.09 &  \citet{ulanowicz86} \\
Charca de Maspalomas Lagoon  &  mgC m$^{-2}$ d$^{-1}$  & 21 & 0.12 & 1,486,230 & 6,010,331 & 0.18 &  \citet{almunia99} \\
Northern Benguela Upwelling  &  mgC m$^{-2}$ d$^{-1}$  & 24 & 0.21 & 2,282 & 6,609 & 0.05 &  \citet{heymans00} \\
Swartkops Estuary & mgC m$^{-2}$ d$^{-1}$ & 25 & 0.17 & 2,860 & 8,950 & 0.27 &  \citet{scharler05} \\
Sundays Estuary & mgC m$^{-2}$ d$^{-1}$ & 25 & 0.16 & 4,442 & 11,940 & 0.22 &  \citet{scharler05} \\
Kromme Estuary & mgC m$^{-2}$ d$^{-1}$ & 25 & 0.16 & 2,571 & 11,088 & 0.38 &  \citet{scharler05} \\
Neuse Estuary (early summer 1997)  &  mgC m$^{-2}$ d$^{-1}$  & 30 & 0.09 & 4,385 & 13,828 & 0.12 &  \citet{baird04} \\
Neuse Estuary (late summer 1997)  &  mgC m$^{-2}$ d$^{-1}$  & 30 & 0.11 & 4,640 & 13,036 & 0.13 &  \citet{baird04} \\
Neuse Estuary (early summer 1998)  &  mgC m$^{-2}$ d$^{-1}$  & 30 & 0.09 & 4,569 & 14,025 & 0.12 &  \citet{baird04} \\
Neuse Estuary (late summer 1998)  &  mgC m$^{-2}$ d$^{-1}$  & 30 & 0.1 & 5,641 & 15,032 & 0.11 &  \citet{baird04} \\
Gulf of Maine  &  g ww m$^{-2}$ yr$^{-1}$  & 31 & 0.35 & 5,054 & 18,382 & 0.15 &   \citet{link08} \\
Georges Bank  &  g ww m$^{-2}$ yr$^{-1}$  & 31 & 0.35 & 4,381 & 16,890 & 0.18 &  \citet{link08} \\
Middle Atlantic Bight  &  g ww m$^{-2}$ yr$^{-1}$  & 32 & 0.37 & 4,869 & 17,917 & 0.18 &  \citet{link08} \\
Narragansett Bay  &  mgC m$^{-2}$ yr$^{-1}$  & 32 & 0.15 & 693,846 & 3,917,246 & 0.51 &  \citet{monaco97} \\
Southern New England Bight  &  g ww m$^{-2}$ yr$^{-1}$  & 33 & 0.35 & 4,718 & 17,597 & 0.16 &  \citet{link08} \\
Chesapeake Bay   &  mgC m$^{-2}$ yr$^{-1}$  & 36 & 0.09 & 888,791 & 3,227,453 & 0.19 &  \citet{baird89} \\
St. Marks Seagrass, site 1 (Jan.)  &  mgC m$^{-2}$ d$^{-1}$  & 51 & 0.08 & 515 & 1,316 & 0.13 &  \citet{baird98} \\
St. Marks Seagrass, site 1 (Feb.)  &  mgC m$^{-2}$ d$^{-1}$  & 51 & 0.08 & 602 & 1,591 & 0.11 &  \citet{baird98} \\
St. Marks Seagrass, site 2 (Jan.)  &  mgC m$^{-2}$ d$^{-1}$  & 51 & 0.07 & 603 & 1,383 & 0.09 &  \citet{baird98} \\
St. Marks Seagrass, site 2 (Feb.)  &  mgC m$^{-2}$ d$^{-1}$  & 51 & 0.08 & 801 & 1,921 & 0.08 &  \citet{baird98}\\
St. Marks Seagrass, site 3 (Jan.)  &  mgC m$^{-2}$ d$^{-1}$  & 51 & 0.05 & 7,809 & 12,651 & 0.01 & \citet{baird98}\\
St. Marks Seagrass, site 4 (Feb.)  &  mgC m$^{-2}$ d$^{-1}$  & 51 & 0.08 & 1,433 & 2,865 & 0.04 &  \citet{baird98}\\
Sylt R{\o}m{\o} Bight  &  mgC m$^{-2}$ d$^{-1}$  & 59 & 0.08 & 683,448 & 1,781,029 & 0.09 &  \citet{baird04_sylt} \\
Graminoids (wet)  &  gC m$^{-2}$ yr$^{-1}$  & 66 & 0.18 & 6,272 & 13,677 & 0.02 &  \citet{ulanowicz00_graminoids} \\
Graminoids (dry)  &  gC m$^{-2}$ yr$^{-1}$  & 66 & 0.18 & 3,473 & 7,520 & 0.04 &   \citet{ulanowicz00_graminoids} \\
Cypress (wet)  &  gC m$^{-2}$ yr$^{-1}$  & 68 & 0.12 & 1,419 & 2,572 & 0.04 &  \citet{ulanowicz97_cypress} \\
Cypress (dry)  &  gC m$^{-2}$ yr$^{-1}$  & 68 & 0.12 & 1,036 & 1,919 & 0.04 &  \citet{ulanowicz97_cypress}\\
Lake Oneida (pre-ZM)  &  gC m$^{-2}$ yr$^{-1}$  & 74 & 0.22 & 1,035 & 1,698 & 0.00 &  \citet{miehls09_oneida} \\
Lake Quinte (pre-ZM)  &  gC m$^{-2}$ yr$^{-1}$  & 74 & 0.21 & 989 & 1,518 & 0.00 &   \citet{miehls09_quinte} \\
Lake Oneida (post-ZM)  &  gC m$^{-2}$ yr$^{-1}$  & 76 & 0.22 & 811 & 1,463 & 0.00 &  \citet{miehls09_oneida} \\
Lake Quinte (post-ZM)  &  gC m$^{-2}$ yr$^{-1}$  & 80 & 0.21 & 1,163 & 2,108 & 0.01 &   \citet{miehls09_quinte} \\
Mangroves (wet)  &  gC m$^{-2}$ yr$^{-1}$  & 94 & 0.15 & 1,532 & 3,266 & 0.10 &  \citet{ulanowicz99_mangrove} \\
Mangroves (dry)  &  gC m$^{-2}$ yr$^{-1}$  & 94 & 0.15 & 1,531 & 3,272 & 0.10 &  \citet{ulanowicz99_mangrove} \\
Florida Bay (wet)  &  mgC m$^{-2}$ yr$^{-1}$  & 125 & 0.12 & 739 & 2,721 & 0.14 &  \citet{ulanowicz98_fb} \\
Florida Bay (dry)  &  mgC m$^{-2}$ yr$^{-1}$  & 125 & 0.13 & 548 & 1,779 & 0.08 &  \citet{ulanowicz98_fb} \\ [-1.5ex]
\end{tabular}
\end{footnotesize}
\end{center}
\tableline
\begin{scriptsize}
$^\dagger$ $n$ is the number of nodes in the network model, $C=L/n^2$
is the model connectance when $L$ is the number of direct links or
energy--matter transfers, $TST=\sum\sum{f_{ij}}+\sum{z_i}$ is the total
system throughflow, and $FCI$ is the Finn Cycling Index \citep{finn80}.  
\end{scriptsize}
\end{table*}

\begin{center}
\begin{table*}
  \caption{Dominant ecosystem components as identified by throughflow
    centrality with \colorbox{myGreen}{primary producers} labeled with
    a green box, \colorbox{myBrown}{\color{white}{dead organic matter}} colored in a brown box with white
    letters, and \colorbox{myPink}{bacteria} in a pink box. These are the model nodes required to generate 50\% of total
    system throughflow (N$_{50}$).} \label{tab:N50}
  \tableline 
\begin{tiny}
\begin{tabular}{l l l l l }
Model & $T_1$ & $T_2$ & $T_3$ & $T_4$ \\ 
\hline \\[-1.5ex]
Bothnian Bay & \colorbox{myBrown}{\color{white}{DOM}} & \colorbox{myPink}{Bacteria} & \colorbox{myBrown}{\color{white}{Sediment C}} & \colorbox{myGreen}{Pelagic Producers} \\
Bothnian Sea & Macrofauna & \colorbox{myBrown}{\color{white}{Sediment Carbon}} & \colorbox{myGreen}{Pelagic Producers} \\
Ythan Estuary & \colorbox{myBrown}{\color{white}{Nutrient Pool}} & \colorbox{myBrown}{\color{white}{Suspended POC}} & \colorbox{myGreen}{Benthic Macrophytes} \\
Sundarban Mangrove (virgin) & \colorbox{myBrown}{\color{white}{Detritus}} & \colorbox{myGreen}{Macrophytes} \\
Sundarban Mangrove (reclaimed) & \colorbox{myBrown}{\color{white}{Detritus}} & \colorbox{myGreen}{Macrophytes} & \colorbox{myGreen}{Benthic algae} \\
Baltic Sea & \colorbox{myGreen}{Pelagic Production} & Mesozooplankton & \colorbox{myBrown}{\color{white}{Suspended POC}} \\
Ems Estuary & \colorbox{myBrown}{\color{white}{Sediment POC}} & \colorbox{myGreen}{Pelagic Producers} & \colorbox{myGreen}{Benthic Producers} \\
Southern Benguela Upwelling & \colorbox{myBrown}{\color{white}{Suspended POC}} & \colorbox{myGreen}{Phytoplankton} \\
Peruvian Upwelling & \colorbox{myGreen}{Pelagic Producers} & Mesozooplankton \\
Crystal River (control) & \colorbox{myGreen}{Macrophytes} & \colorbox{myBrown}{\color{white}{Detritus}} \\
Crystal River (thermal) & \colorbox{myGreen}{Macrophytes} & \colorbox{myBrown}{\color{white}{Detritus}} \\
Charca de Maspalomas Lagoon & \colorbox{myBrown}{\color{white}{Sedimented POC}} & Mesozooplankton & Benthic Deposit Feeders & \colorbox{myGreen}{Cyanobacteria} \\
Northern Benguela Upwelling & \colorbox{myBrown}{\color{white}{POC}} & \colorbox{myBrown}{\color{white}{DOC}} & \colorbox{myPink}{Bacteria} \\
Swartkops Estuary & \colorbox{myBrown}{\color{white}{Sediment POC}} & \colorbox{myPink}{Sediment Bacteria} \\
Sundays Estuary & \colorbox{myBrown}{\color{white}{Sediment POC}} & \colorbox{myPink}{Sediment Bacteria} & \colorbox{myGreen}{Phytoplankton} \\
Kromme Estuary & \colorbox{myBrown}{\color{white}{Sediment POC}} & \colorbox{myPink}{Sediment Bacteria} \\
Neuse Estuary (early summer 1997) & \colorbox{myPink}{Free Living Bacteria} & \colorbox{myBrown}{\color{white}{DOC}} & \colorbox{myBrown}{\color{white}{Sediment POC}} & \colorbox{myPink}{Sediment Bacteria} \\
Neuse Estuary (late summer 1997)  & \colorbox{myBrown}{\color{white}{DOC}} & \colorbox{myPink}{Free Living Bacteria} & \colorbox{myBrown}{\color{white}{Sediment POC}} \\
Neuse Estuary (early summer 1998) & \colorbox{myPink}{Free Living Bacteria} & \colorbox{myBrown}{\color{white}{DOC}} & \colorbox{myBrown}{\color{white}{Sediment POC}} \\
Neuse Estuary (late summer 1998) & \colorbox{myBrown}{\color{white}{DOC}} & \colorbox{myPink}{Free Living Bacteria} & \colorbox{myGreen}{Phytoplankton} \\
Gulf of Maine & \colorbox{myGreen}{Phytoplankton-Primary} & Large Copepods & \colorbox{myBrown}{\color{white}{Detritus--POC}} \\
Georges Bank & \colorbox{myGreen}{Phytoplankton-Primary} & \colorbox{myBrown}{\color{white}{Detritus--POC}} & \colorbox{myPink}{Bacteria} \\
Middle Atlantic Bight & \colorbox{myGreen}{Phytoplankton-Primary} & \colorbox{myBrown}{\color{white}{Detritus--POC}} & \colorbox{myPink}{Bacteria} \\
Narragansett Bay & \colorbox{myBrown}{\color{white}{Detritus}} & \colorbox{myPink}{Sediment POC Bacteria} \\
Southern New England Bight & \colorbox{myGreen}{Phytoplankton-Primary} & \colorbox{myBrown}{\color{white}{Detritus--POC}} & \colorbox{myPink}{Bacteria} \\
Chesapeake Bay  & \colorbox{myBrown}{\color{white}{Sediment Particulate Carbon}} & \colorbox{myPink}{Bacteria in Sediment POC} & \colorbox{myGreen}{Phytoplankton} \\
St.\ Marks Seagrass, site 1 (Jan.) & \colorbox{myPink}{Benthic Bacteria} & \colorbox{myGreen}{Micro-epiphytes} & \colorbox{myBrown}{\color{white}{Sediment POC}} \\
St.\ Marks Seagrass, site 1 (Feb.) & \colorbox{myPink}{Benthic Bacteria} & \colorbox{myBrown}{\color{white}{Sediment POC}} & \colorbox{myGreen}{Benthic algae} & Meiofauna \\
St.\ Marks Seagrass, site 2 (Jan.) & \colorbox{myGreen}{Micro-epiphytes} & \colorbox{myBrown}{\color{white}{Sediment POC}} \\
St.\ Marks Seagrass, site 2 (Feb.) & \colorbox{myBrown}{\color{white}{Sediment POC}} & \colorbox{myGreen}{Benthic algae} & \colorbox{myPink}{Benthic Bacteria} \\
St.\ Marks Seagrass, site 3 (Jan.) & \colorbox{myGreen}{Micro-epiphytes} \\
St.\ Marks Seagrass, site 4 (Feb.) & Pinfish & \colorbox{myBrown}{\color{white}{Sediment POC}} \\
Sylt-R{\o}m{\o} Bight & \colorbox{myBrown}{\color{white}{Sediment POC}} & \colorbox{myGreen}{Microphytobenthos} & \colorbox{myGreen}{Phytoplankton} \\
Everglade Graminoids (wet) & \colorbox{myBrown}{\color{white}{Sediment Carbon}} & \colorbox{myGreen}{Periphyton} & \colorbox{myBrown}{\color{white}{Refractory Detritus}} \\
Everglade Graminoids (dry) & \colorbox{myGreen}{Periphyton} & \colorbox{myBrown}{\color{white}{Sediment Carbon}} \\
Cypress (wet) & \colorbox{myBrown}{\color{white}{Refractory Detritus}} & \colorbox{myGreen}{Cypress} & \colorbox{myPink}{Living Sediment} & \colorbox{myBrown}{\color{white}{Liable Detritus}} \\
Cypress (dry) & \colorbox{myBrown}{\color{white}{Refractory Detritus}} & \colorbox{myPink}{Living sediment} & \colorbox{myGreen}{Understory} & \colorbox{myBrown}{\color{white}{Liable Detritus}} \\
Lake Oneida (pre-ZM) & \colorbox{myBrown}{\color{white}{Pelagic Detritus}} & \colorbox{myGreen}{Diatoms} & \colorbox{myGreen}{Blue-myGreen Algae} & \colorbox{myGreen}{Epiphytes} \\
Lake Quinte (pre-ZM) & \colorbox{myBrown}{\color{white}{Pelagic Detritus}} & \colorbox{myGreen}{Diatoms} \\
Lake Oneida (post-ZM) & \colorbox{myGreen}{Diatoms} & \colorbox{myGreen}{Epiphytes} & \colorbox{myBrown}{\color{white}{Pelagic Detritus}} & \colorbox{myGreen}{Blue-myGreen Algae} \\
Lake Quinte (post-ZM) & Zebra Mussels & \colorbox{myGreen}{Diatoms} \\
Mangroves (wet) & \colorbox{myBrown}{\color{white}{Carbon in Sediment}} & \colorbox{myGreen}{Leaf} & \colorbox{myGreen}{Other Primary Producers} \\
Mangroves (dry) & \colorbox{myBrown}{\color{white}{Carbon in Sediment}} & \colorbox{myGreen}{Leaf} & \colorbox{myGreen}{Other Primary Producers} \\
Florida Bay (wet) & \colorbox{myBrown}{\color{white}{Benthic POC}} & \colorbox{myBrown}{\color{white}{Water POC}} & \colorbox{myGreen}{Water Flagellates} & \colorbox{myGreen}{Thalassia} \\
Florida Bay (dry) & \colorbox{myBrown}{\color{white}{Benthic POC}} & \colorbox{myBrown}{\color{white}{Water POC}} & \colorbox{myGreen}{Thalassia} & \colorbox{myBrown}{\color{white}{DOC}} \\
\end{tabular}
\tableline
\end{tiny}
\end{table*}
\end{center}

\clearpage


\end{document}